\documentclass[conference]{IEEEtran}
\IEEEoverridecommandlockouts
\usepackage{cite}
\usepackage{amsmath,amssymb,amsfonts}
\usepackage{algorithmic}
\usepackage{graphicx}
\usepackage{textcomp}
\usepackage[caption=false,font=footnotesize]{subfig}
\usepackage{wrapfig}
\usepackage{listings}
\lstdefinestyle{ieeecode}{
    backgroundcolor=\color{white},
    basicstyle=\ttfamily\footnotesize,
    columns=fullflexible,
    breaklines=true,
    frame=single,
    captionpos=b,
    language=Java, 
    numbers=none,
}
\usepackage[table,xcdraw]{xcolor} 
\usepackage{xcolor}
\def\BibTeX{{\rm B\kern-.05em{\sc i\kern-.025em b}\kern-.08em
    T\kern-.1667em\lower.7ex\hbox{E}\kern-.125emX}}
\begin{document}

\title{Enhancing Evacuation Safety: Detecting Post-Nuclear Event Radiation Levels in an Urban Area}

\author{\IEEEauthorblockN{Ellis Duncalfe}
\IEEEauthorblockA{\textit{School of Computer Science} \\
\textit{University of Nottingham}\\
Nottingham, United Kingdom \\
psyed4@nottingham.ac.uk}
\and
\IEEEauthorblockN{Milena Radenkovic}
\IEEEauthorblockA{\textit{School of Computer Science} \\
\textit{University of Nottingham}\\
Nottingham, United Kingdom \\
milena.radenkovic@nottingham.ac.uk}
}

\maketitle

\begin{abstract}
The detonation of an improvised nuclear device (IND) in an urban area would cause catastrophic damage, followed by hazardous radioactive fallout. Timely dissemination of radiation data is crucial for evacuation and casualty reduction. However, conventional communication infrastructure is likely to be severely disrupted. This study designs and builds a pseudorealistic, geospatially and temporally dynamic post-nuclear event (PNE) scenario using the Opportunistic Network Environment (ONE) simulator. It integrates radiation sensing by emergency responders, unmanned aerial vehicles (UAVs), and civilian devices as dynamic nodes within Delay-Tolerant Networks (DTNs). The performance of two DTN routing protocols, Epidemic and PRoPHET, was evaluated across multiple PNE phases. Both protocols achieve high message delivery rates, with PRoPHET exhibiting lower network overhead but higher latency. Findings demonstrate the potential of DTN-based solutions to support emergency response and evacuation safety by ensuring critical radiation data propagation despite severe infrastructure damage.
\end{abstract}


\section{Introduction}

Nuclear attacks on major urban areas pose a significant and growing national security threat. The detonation of an improvised nuclear device (IND) in a city would result in immediate catastrophic effects, including mass casualties from the blast, followed by hazardous radioactive fallout. Fallout occurs when dust and debris excavated by the explosion combine with radioactive fission products, creating a persistent environmental hazard \cite{osti_1119909}. In such an event, the importance of proper sheltering and evacuation strategies cannot be overstated. If timely and well-coordinated, these measures can significantly reduce the number of casualties and limit radiation exposure. These strategies, however, rely on timely and accurate dissemination of critical information, such as radiation level data.

Urban environments present unique challenges in disaster response due to their dense infrastructure and population. Conventional communication networks - such as cellular and fixed-line systems - are likely to be destroyed or overwhelmed with congestion in the event of a nuclear incident, severely limiting their reliability for critical information sharing. While satellite imagery can provide broad-area assessment, they have limited temporal resolution \cite{Smith2024RemoteST} and cannot provide real-time data from specific locations where immediate action is required. 

Delay-Tolerant Networks (DTNs), particularly Opportunistic Networks (OppNets), offer a well-suited alternative communication paradigm for disrupted environments such as this. They utilise a store-carry-forward technique \cite{OpportunisticNetworks} in peer-to-peer mobile nodes, making them resilient to network failures \cite{inbook}.

This paper explores the challenge of enabling real-time radiation data dissemination in a post-nuclear urban environment through the design and construction of a pseudorealistic, geospatially and temporally dynamic scenario. Using the Opportunistic Network Environment (ONE) simulator, we integrate emergency responders, unmanned aerial vehicles (UAVs) and civilian devices as mobile nodes that perform data collection and opportunistic message relaying. The study evaluates the performance of two DTN protocols, Epidemic and PRoPHET, across multiple post-nuclear event (PNE) phases to assess their performance in reliably propagating radioactivity level data that could be used by emergency responders to map the safest evacuation routes for civilians.

\section{Related Work}

\subsection{Opportunistic Networks, Mobile Ad-hoc Networks (MANETs) \& Vehicular Ad-hoc Networks (VANETs)}

Opportunistic Networks (OppNets) are a type of Delay-Tolerant Network (DTN) designed for environments with intermittent connectivity. They are based on Mobile Ad-Hoc Networks (MANETs) -  group of autonomous mobile nodes that communicate using a shared wireless multi-hop channel, requiring no infrastructure. In this architecture, every computer (node) is a router as well as an end host \cite{mirza2018introduction}, forwarding data to other devices in the network. 

Routing in MANETs can be proactive, reactive, or hybrid, all of which assume a relatively stable path exists between sender and receiver \cite{OpportunisticNetworks}. However, in real-world disaster scenarios, maintaining continuous end-to-end paths is often infeasible due to node mobility and wireless link instability, as well as infrastructure loss.

Vehicular Ad-hoc Networks (VANETs) are a type of MANET adapted for vehicle to vehicle (V2V), vehicle to roadside unit/infrastructure (RSU/V2I), or vehicle to anything (V2X) communication. VANETs differ from MANETs mainly in their predictable mobility patterns (due to movement along set routes like roads or tracks) and larger communication ranges, which accommodate high-speed vehicle movements and increases node connection time \cite{ALSULTAN2014380}. 

OppNets address the limitations caused by intermittent connectivity in traditional ad-hoc networks by using a  store-carry-forward strategy. Data is temporarily stored in a buffer at intermediate nodes and carried until a suitable next hop is encountered to forward to, allowing messages to propagate even without end-to-end connections \cite{OpportunisticNetworks}.

Conventional communication infrastructure is likely to be destroyed or overwhelmed in a nuclear disaster scenario, as well as nodes experiencing intermittent connectivity, making OppNets a highly suited alternative. Smartphones of civilians, emergency responders, UAVs, and stationary sensors can act as opportunistic nodes that collect and disseminate critical data such as radiation levels or survivor locations. UAVs and ground vehicles are particularly valuable in such scenarios due to their ability to traverse affected areas that others could not, and provide flexible communication capabilities. Prior work has demonstrated the effectiveness of using heterogeneous networks of drones and vehicles to provide real-time data collection and adaptive communication services in disconnected environments \cite{realtime}. These platforms, such as MODiToNeS \cite{moditones}, have shown reliability and flexibility in environments with physical obstructions and intermittent connectivity, which aligns closely with the needs of post-disaster OppNets.

\begin{figure}
    \centering
    \includegraphics[width=0.45\textwidth]{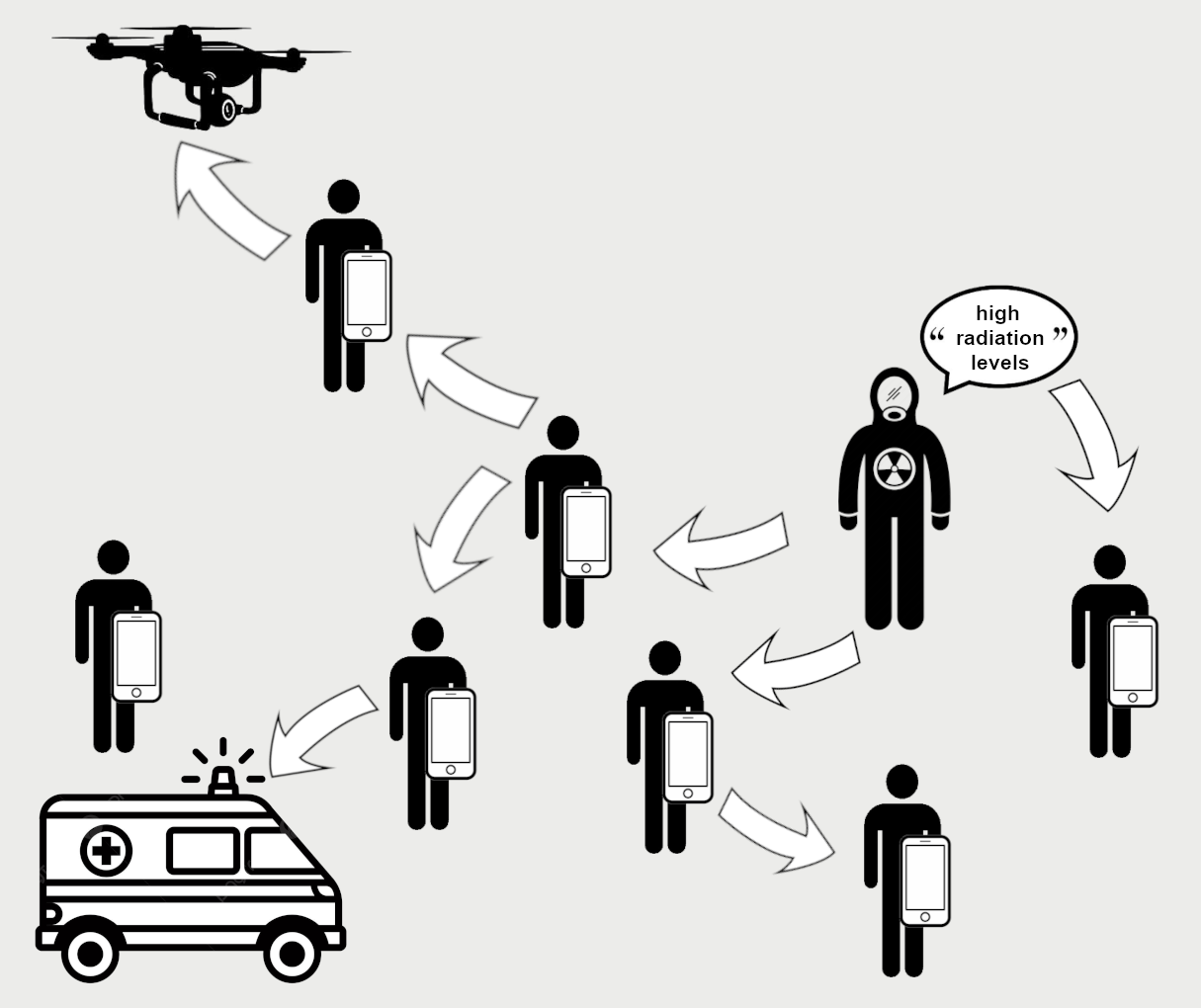}
    \caption{Dissemination of radiation level data via multi-hops from a first responder to UAV and emergency vehicle through civilian mobile devices.}
    \label{fig:diagram}
\end{figure}

For instance, as nodes move through areas with hazardous radiation levels present, this data can be relayed from isolated zones to command centers, ensuring that decision makers have up-to-date situational awareness, and enabling mapping of the fallout zone. As illustrated in Figure \ref{fig:diagram}, a first responder may transmit radiation measurements to multiple nearby devices (multicast), which store and forward the data to other nodes until it reaches the intended destination host(s) (in this case, a UAV or ambulance). This information can then be redistributed to inform survivors of the safest areas to minimise radiation exposure. 

Several lightweight platforms have been developed to support such infrastructure-less communication. One such system is RasPiPCloud, a Raspberry Pi-based personal cloud that supports data storage, management, and sharing through an architecture that can be used in a disconnection-tolerant manner \cite{raspipcloud}. As it's designed for portability and low-cost deployment, RasPiPCloud demonstrates how personal cloud infrastructure can be extended to remote or under-resourced environments, compatible with emergency deployment scenarios.

\subsection{Delay-Tolerant Network (DTN) Routing Protocols}

There are many routing protocols for DTNs, each with different approaches to handling the challenges
of intermittent connectivity and absence of end-to-end paths. For this emergency scenario, Epidemic and
PRoPHET were chosen to compare due to their relevance in balancing delivery probability and resource
efficiency, but contrasting approaches in doing so.

\subsubsection{Epidemic}

Epidemic routing employs a flooding approach that prioritises maximising message delivery probability by replicating messages across all encountered nodes \cite{epidemic}. It assumes inexact knowledge
of the location of nodes throughout the system, and these minimal assumptions make it well-suited for
unpredictable environments such as post-disaster scenarios.

When two nodes meet, they exchange summary vectors to determine if either of them hold messages the other does not already have, and if so they create a copy to share, synchronising their buffers. This ensures wide dissemination of each message regardless of network dynamics.

While the protocol offers high delivery success and low latency in highly unpredictable environments, such as post-disaster zones, it incurs high overhead due to uncontrolled message replication. This can result in significant resource consumption that reduces scalability in dense networks.

\subsubsection{PRoPHET}

The PRoPHET (Probabilistic Routing Protocol using History of Encounters and Transitivity) protocol introduces a more efficient alternative based on delivery predictability metrics \cite{prophet}. Each node maintains a table estimating the likelihood of successful message delivery to other nodes, based on how often they interact, and patterns in which they do so. 

Messages are forwarded selectively to nodes with higher delivery probabilities, reducing redundant transmissions. This results in reduced overhead and converses resources in networks with semi-predictable mobility patterns. However, in scenarios where movement is highly random, selective forwarding may result in missed delivery opportunities. 

\subsection{The Opportunistic Network Environment (ONE) Simulator}

The Opportunistic Network Environment (ONE) Simulator \cite{TheONESimulator} is a tool for modelling and evaluating DTNs. It enables simulation of node mobility, inter-node contacts, message routing, and data dissemination in intermittently connected networks. It supports integration of real-world maps with customisable mobility models for nodes, enabling accurate representation of real-life scenarios.

The different entities that could be present in the aftermath of a nuclear event - such as survivors, emergency responders, UAVs and radiation zones - can be emulated by defining their respective roles, movement behaviours, and constraints. Parameters such as node density, buffer size, transmission range, and mobility patterns can be configured to simulate network state under specific conditions.

The simulator produces key performance metrics such as delivery probability, message latency, and overhead ratio, which are essential in evaluating the performance of routing protocols. These insights could help prepare strategies and assist in designing robust DTN deployments capable of supporting data dissemination in post-nuclear disaster environments.

\section{Modelling Geospatial and Temporal Changes in Disaster Zones}

This study simulates the aftermath of a nuclear explosion in the center of a city, with the goal of evaluating
the reliability, efficiency and speed of message dissemination of two DTN protocols. These metrics will help
to determine whether essential data reaches the necessary destination nodes to be able to create informed
decisions and further distribute information to emergency services and the public, such as evacuation routes
or hazardous areas.

The scenario involves various types of nodes representing real-world entities with distinct roles and capabilities:
\begin{itemize}
    \item \textbf{Pedestrian Civilians:} Represent individuals on foot. They can relay messages but are unable to detect radiation levels directly. It is assumed that the node is a smartphone device or similar with no dosimeter or similar radiation detection functionality. Their movement is constrained to roads and pedestrian pathways.
    \item \textbf{Civilian Vehicles:} Represent private or public transport vehicles. They can relay messages but are unable to detect radiation levels directly. Their movement is restricted to roads, avoiding areas that have been mapped as 'destroyed' by the blast zone.
    \item \textbf{Emergency Responder Vehicles:} Represent vehicles used by first responders. Equipped with a dosimeter, Geiger counter, or similar, they can measure radiation levels and generate messages to share the information. Their movement is restricted to roads, avoiding areas that have been mapped as 'destroyed' by the blast zone.
\end{itemize}

\begin{itemize}
    \item \textbf{Unmanned Aerial Vehicles (UAVs):} Play a crucial role by flying over the affected area, detecting radiation levels and assessing damage with the least risk, disseminating the information to nodes they encounter. They are not constrained by the map at all as they are airborne.
\end{itemize}

Emergency responder vehicles and UAVs act as the primary sensors, detecting radiation and broadcasting the information. Civilians are able to relay these messages through the network. This ensures that accurate and up-to-date information about radiation levels spreads across the network, even with limited infrastructure.

To represent the evolving conditions and responses over time, the simulations are divided into four distinct post-nuclear event (PNE) phases, each reflecting a different period after the initial detonation:
\begin{itemize}
    \item \textbf{Phase 1 (0-12 hours):} This critical initial phase simulates the immediate aftermath of the explosion. Communication infrastructure is highly unstable, and although the priority is to rapidly collect and disseminate radiation data \cite{nuclearplanningguidance}, official advice for those affected is to find immediate shelter and remain there for 2 weeks if possible, or a minimum of 48 hours to reduce risk of life-threatening radiation exposure \cite{Dillon2014DeterminingOF}. This is represented by a medium number of pedestrian nodes moving through the affected zones, as some may have not found shelter or may be unaware of the danger of remaining outside. People who are taking shelter are not represented in the simulation, as they are likely to be underground or otherwise unreachable. The deployment of a singular UAV to begin data collection is simulated. The majority of node movement is limited to outside of the affected zones, as most people will not head towards them.
    \item \textbf{Phase 2 (12-48 hours):} This phase reflects the period when more emergency services are deployed, along with another UAV to assess damage. Within the zone, the number of civilian nodes drops, as most will be remaining in their shelter for as long as possible, or managed to leave the zone. Additionally, civilians external to the zones also drop in number, as many would choose to evacuate or stay at home in this event. Emergency responders are not set to enter the hazardous zones yet, as it is unlikely that there is equipment or enough information to allow them to safely begin recovery measures. There will also be a more limited number of people willing to enter areas affected by radiation \cite{Dallas2017ReadinessFR}. However, they are able to move around the edge of the outer zone, and in doing so may occasionally pick up radiation readings that can be shared. The UAVs continue to fly over and collect information.
    \item \textbf{Phase 3 (48 hours-7 days):} In this phase, efforts to stabilize the situation intensify. Some emergency responders are now able to enter the affected zones to begin rescue of civilians and collect radiation data. More civilians exit their shelters as they run out of supplies, so are contactable nodes. Another UAV is used to collect further data as the radiation levels deplete over time.
    \item \textbf{Phase 4 (7-14 days):} This phase represents a later recovery effort. Radiation levels drop and are much less life-threatening \cite{osti_966550}, so more civilians leave their shelters to evacuate. This also means more emergency responders can enter the zones and recover victims, as there is much less threat to their lives. Another UAV is deployed. Increased nodes in the affected zones mean more messages, but also a more connected network.
\end{itemize}

\subsection{Radiation Zone Construction}

\begin{figure}
    \centering
    \includegraphics[width=0.45\textwidth]{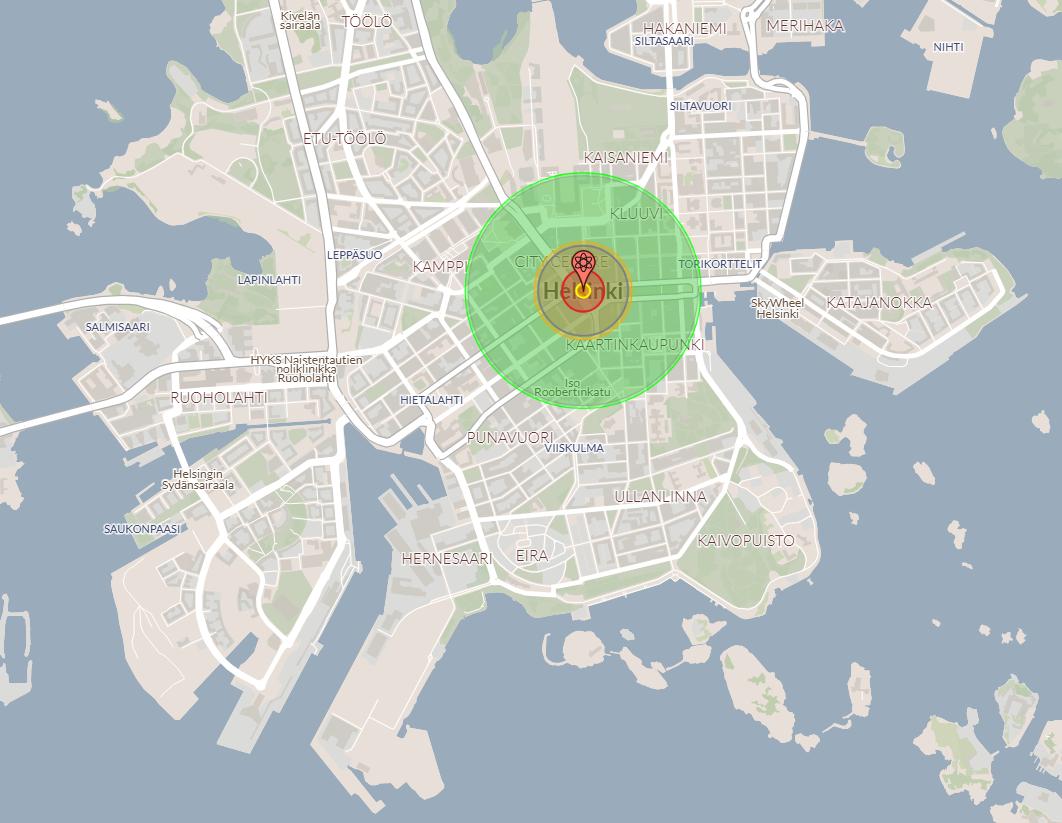}
    \caption{Modelled affected areas of a 100t crude nuclear terrorist weapon detonated in the heart of Helsinki.}
    \label{fig:nukemap}
\end{figure}

The urban city used for this experiment environment is Helsinki, as the ONE simulator already provided pre-processed WKT (Well-Known Text) files for it. These files contain spatial data describing the road networks within the map. An online tool for modelling nuclear explosion effects called NukeMap \cite{NUKEMAP} was used to simulate the aftermath of a nuclear terrorist attack. Figure \ref{fig:nukemap} shows the results of this simulation. The different radii represent the fireball radius, heavy blast damage, moderate blast damage, thermal radiation, light blast damage and radiation zones. Radiation levels increase as you approach the blast point.

\begin{figure}[!b]
    \centering
    \subfloat[]{\includegraphics[width=0.32\linewidth]{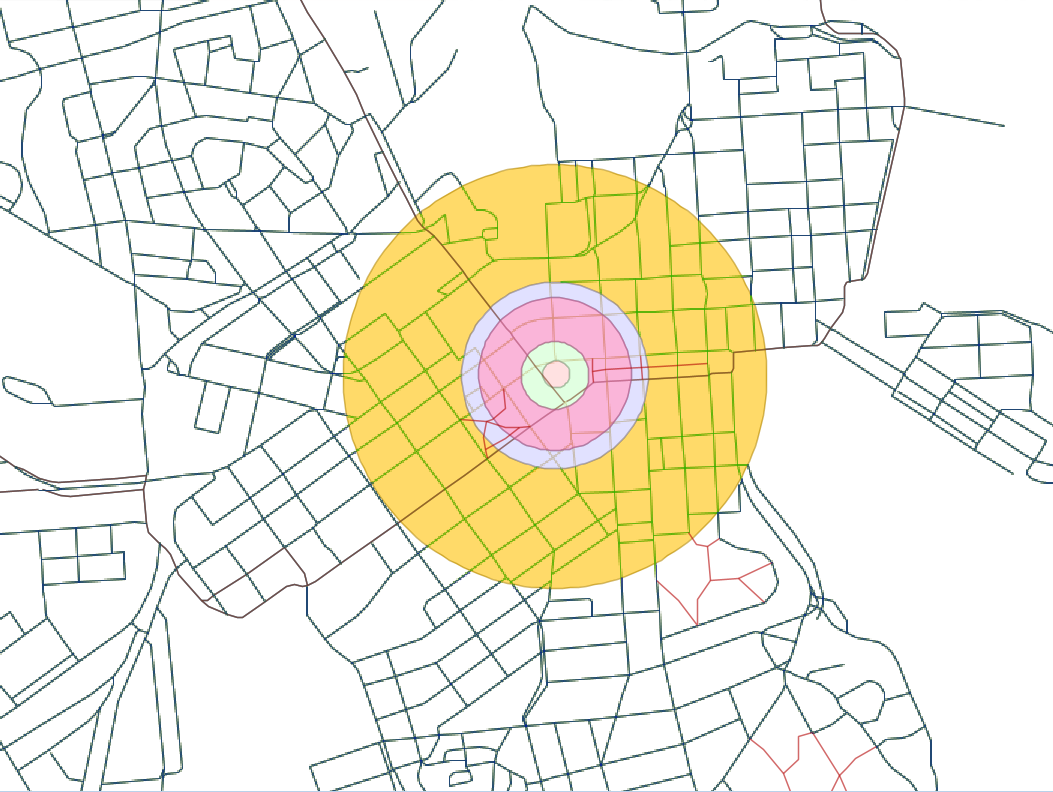}} 
    \hfill
    \subfloat[]{\includegraphics[width=0.32\linewidth]{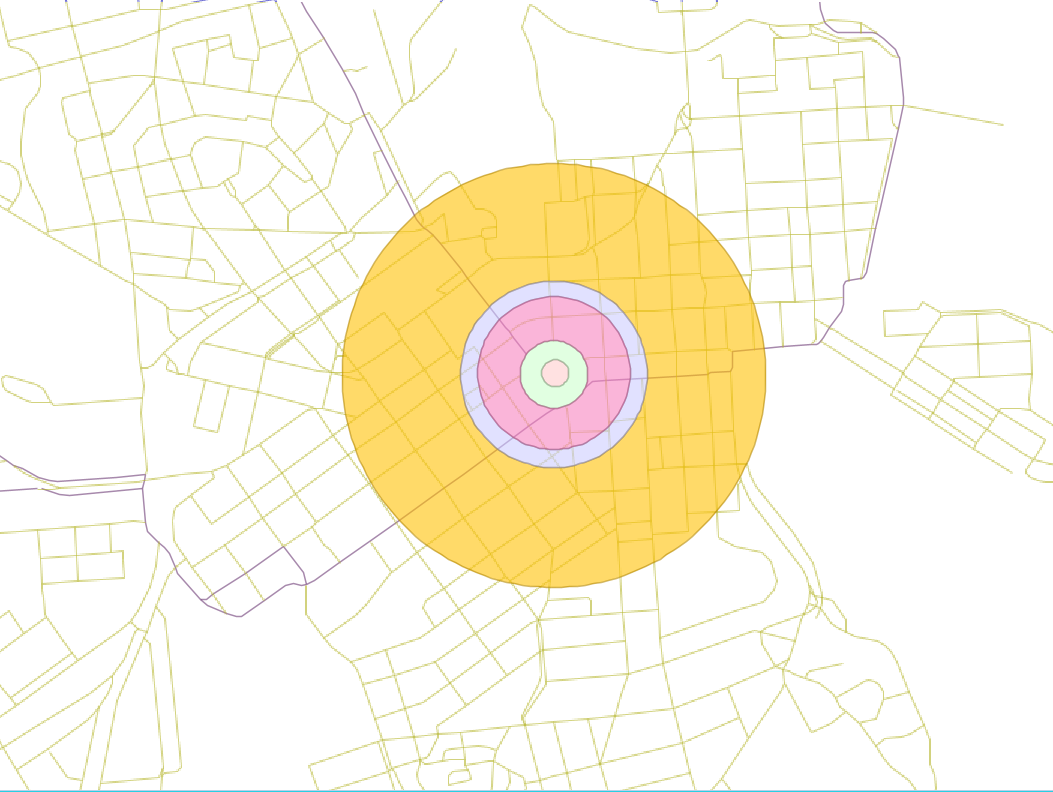}} 
    \hfill
    \subfloat[]{\includegraphics[width=0.32\linewidth]{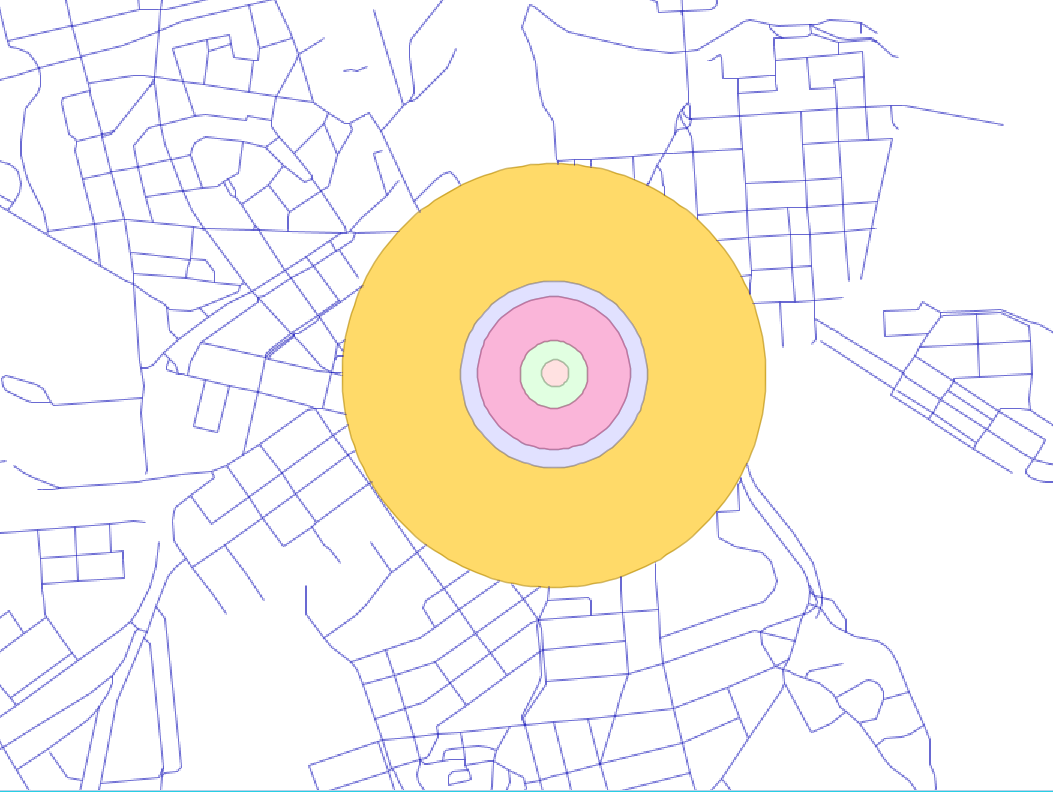}}
    \caption{(a) All road data (b) Roads excluding zones 1 and 2 (c) Roads excluding all zones}
    \label{fig:wktfiles}
\end{figure}

\begin{figure}
    \centering
    \includegraphics[width=0.45\textwidth]{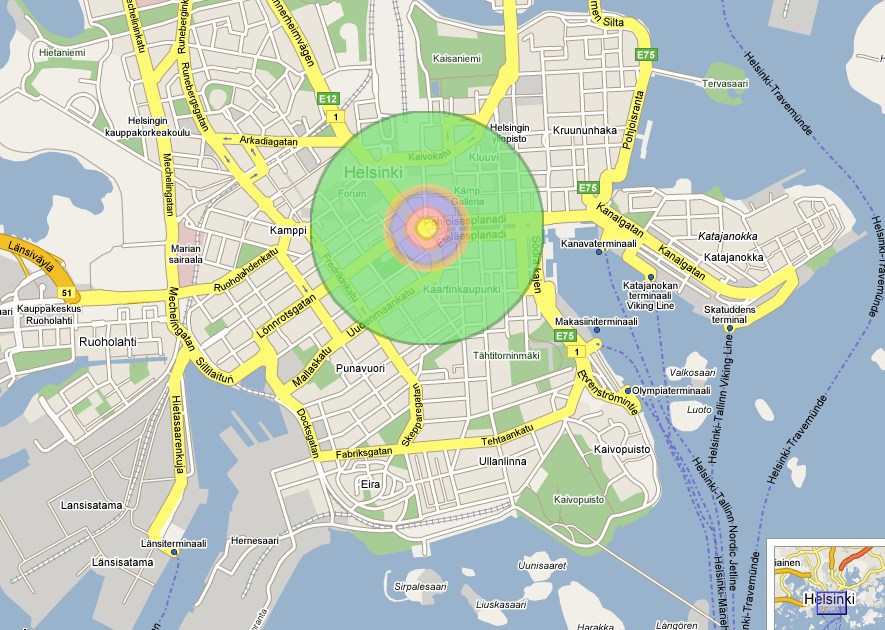}
    \caption{The map used in the simulator.}
    \label{fig:simulatormap}
\end{figure}

In order to integrate the blast zones into the simulation, the WKT files of the Helsinki map were modified using OpenJUMP \cite{OpenJump}, a GIS (Geographic Information System) tool. This involved deriving radii from the NukeMap and converting them into relative distances on the Helsinki map, drawing concentric circles to reflect them. Figure \ref{fig:wktfiles} (a) shows the original WKT files that map the road networks throughout Helsinki with the estimated affected zones as an overlay. Removing zones 1 and 2 (the innermost zones) produces Figure \ref{fig:wktfiles} (b), which is used for node movement within the zones. This replicates the fact that the two most inner points would be completely destroyed and unnavigable. For node movement external to the zones, the map shown in Figure \ref{fig:wktfiles} (c) is used, as this removes all roads within the areas, preventing nodes from spawning or moving within this area. These modifications allow the simulator to enforce realistic movement constraints on nodes, depending on their roles. This is configured in the ONE simulator by importing the WKT files as different maps and assigning each node the appropriate 'okMap' properties. Figure \ref{fig:simulatormap} shows the map used in ONE, which was created to replicate the NukeMap model. This gives clear indication of what zone a node is in visually.

\subsection{Designing Entities in ONE}

The ONE simulator configuration file can be set to represent the above node groups and phases, and these settings are displayed in Table \ref{tab:table1}. Due to limitations in computational power, the number of nodes has been scaled down.

Bluetooth is a common short-range wireless communication protocol \cite{McDermottWells2005WhatIB}, and has been chosen as the interface for civilians and their vehicles as they will typically lack access to long-range technologies. Modern devices will all have this functionality so it is a good representation of the nodes in an urban environment. Additionally, the emergency responders will also have access to this protocol. However, UAVs will only have access to a longer range highspeed interface, as they will be travelling at speed and from great heights. This would most likely be Wi-Fi with a range of 100 meters \cite{Gu2015AirborneWN}. Emergency responders have been allowed access to this interface also as they will be communicating with the UAVs.

The buffer sizes were assigned based on the expected role and hardware capacity of each node time. Civilians are assumed to have limited-capacity devices, like smartphones, which can only store a small number of messages. Their vehicles are likely to have slightly higher storage. For emergency responders, their vehicles are equipped with more advanced systems capable of storing and processing larger amounts of data. A higher buffer size ensures they can handle the additional responsibility of detecting and broadcasting critical information. The same can be said for the UAVs, and as they may be retrieving more data than just the radiation levels, such as video, their buffer has been set the highest.

Wait times were selected to represent realistic movement and interaction patterns for each node type, changing across different phases. For example, pedestrian civilians have longer wait times to reflect their slower, less frequent movement within the environment. Vehicles have shorter wait times to simulate faster, more frequent movement patterns. UAVs have minimal wait times to reflect their continuous flight and scanning behaviour.

Along with the above, the message TTL (Time To Live) has been set to 5 hours.

\begin{table*}[!t]
\centering
\caption{Different simulation scenario configurations for different phases of the emergency.}
\resizebox{\textwidth}{!}{%
\begin{tabular}{|c|l|c|c|c|c|c|c|c|}
\hline
\rowcolor[HTML]{ECF4FF} 
\textbf{Phase} & \multicolumn{1}{c|}{\textbf{Node Group}} & \textbf{Inside Zones} & \textbf{Node Count} & \textbf{Buffer Size} & \textbf{Interfaces} & \textbf{Movement Model} & \textbf{Movement Speed (m/s)} & \textbf{Wait Time (s)} \\ 
\hline

\rowcolor[HTML]{F3F3F3}
{(1) 0 – 12 hrs} & Civilians (Pedestrians) & True & 15 & 256k & Bluetooth & ShortestMapBasedMovement & 0.9, 4 & 60, 43200 \\ \cline{2-9}
\rowcolor[HTML]{F3F3F3}
 & Civilians (Pedestrians) & False & 70 & 256k & Bluetooth & ShortestMapBasedMovement & 0.5, 2.1 & 30, 3600 \\ \cline{2-9}
\rowcolor[HTML]{F3F3F3}
 & Civilians (Vehicles) & True & 5 & 512k & Bluetooth & ShortestMapBasedMovement & 2.7, 13.2 & 30, 300 \\ \cline{2-9}
\rowcolor[HTML]{F3F3F3}
 & Civilians (Vehicles) & False & 30 & 512k & Bluetooth & ShortestMapBasedMovement & 1.5, 7.7 & 60, 300 \\ \cline{2-9}
\rowcolor[HTML]{F3F3F3}
 & Emergency Responders & False & 5 & 2M & Bluetooth, Highspeed & ShortestMapBasedMovement & 2.7, 13.2 & 120, 600 \\ \cline{2-9}
\rowcolor[HTML]{F3F3F3}
 & UAVs & True & 1 & 5M & Highspeed & RandomWaypoint & 10, 25 & 10, 60 \\ 
\hline

\rowcolor[HTML]{FFFFFF}
{(2) 12 – 48 hrs} & Civilians (Pedestrians) & True & 2 & 256k & Bluetooth & ShortestMapBasedMovement & 0.0, 3.5 & 600, 43200 \\ \cline{2-9}
\rowcolor[HTML]{FFFFFF}
 & Civilians (Pedestrians) & False & 50 & 256k & Bluetooth & ShortestMapBasedMovement & 0.5, 2.1 & 30, 3600 \\ \cline{2-9}
\rowcolor[HTML]{FFFFFF}
 & Civilians (Vehicles) & True & 1 & 512k & Bluetooth & ShortestMapBasedMovement & 2.7, 13.2 & 30, 300 \\ \cline{2-9}
\rowcolor[HTML]{FFFFFF}
 & Civilians (Vehicles) & False & 20 & 512k & Bluetooth & ShortestMapBasedMovement & 1.5, 13.2 & 60, 300 \\ \cline{2-9}
\rowcolor[HTML]{FFFFFF}
 & Emergency Responders & False & 10 & 2M & Bluetooth, Highspeed & ShortestMapBasedMovement & 2.7, 15 & 120, 600 \\ \cline{2-9}
\rowcolor[HTML]{FFFFFF}
 & UAVs & True & 2 & 5M & Highspeed & RandomWaypoint & 10, 25 & 10, 60 \\ 
\hline

\rowcolor[HTML]{F3F3F3}
{(3) 48 hrs – 7 days} & Civilians (Pedestrians) & True & 7 & 256k & Bluetooth & ShortestMapBasedMovement & 0.0, 3.5 & 600, 432000 \\ \cline{2-9}
\rowcolor[HTML]{F3F3F3}
 & Civilians (Pedestrians) & False & 40 & 256k & Bluetooth & ShortestMapBasedMovement & 0.5, 2.1 & 30, 3600 \\ \cline{2-9}
\rowcolor[HTML]{F3F3F3}
 & Civilians (Vehicles) & True & 5 & 512k & Bluetooth & ShortestMapBasedMovement & 2.7, 13.2 & 30, 300 \\ \cline{2-9}
\rowcolor[HTML]{F3F3F3}
 & Civilians (Vehicles) & False & 15 & 512k & Bluetooth & ShortestMapBasedMovement & 1.5, 13.2 & 60, 300 \\ \cline{2-9}
\rowcolor[HTML]{F3F3F3}
 & Emergency Responders & True & 2 & 2M & Bluetooth, Highspeed & ShortestMapBasedMovement & 2.7, 7.7 & 30, 900 \\ \cline{2-9}
\rowcolor[HTML]{F3F3F3}
 & Emergency Responders & False & 10 & 2M & Bluetooth, Highspeed & ShortestMapBasedMovement & 2.7, 15 & 120, 600 \\ \cline{2-9}
\rowcolor[HTML]{F3F3F3}
 & UAVs & True & 3 & 5M & Highspeed & RandomWaypoint & 10, 25 & 10, 150 \\ 
\hline

\rowcolor[HTML]{FFFFFF}
{(4) 7 – 14 days} & Civilians (Pedestrians) & True & 30 & 256k & Bluetooth & ShortestMapBasedMovement & 0.0, 3.5 & 600, 432000 \\ \cline{2-9}
\rowcolor[HTML]{FFFFFF}
 & Civilians (Pedestrians) & False & 60 & 256k & Bluetooth & ShortestMapBasedMovement & 0.5, 2.1 & 30, 3600 \\ \cline{2-9}
\rowcolor[HTML]{FFFFFF}
 & Civilians (Vehicles) & True & 15 & 512k & Bluetooth & ShortestMapBasedMovement & 2.7, 13.2 & 30, 300 \\ \cline{2-9}
\rowcolor[HTML]{FFFFFF}
 & Civilians (Vehicles) & False & 30 & 512k & Bluetooth & ShortestMapBasedMovement & 1.5, 13.2 & 60, 300 \\ \cline{2-9}
\rowcolor[HTML]{FFFFFF}
 & Emergency Responders & True & 10 & 2M & Bluetooth, Highspeed & ShortestMapBasedMovement & 2.7, 7.7 & 30, 900 \\ \cline{2-9}
\rowcolor[HTML]{FFFFFF}
 & Emergency Responders & False & 15 & 2M & Bluetooth, Highspeed & ShortestMapBasedMovement & 2.7, 15 & 120, 600 \\ \cline{2-9}
\rowcolor[HTML]{FFFFFF}
 & UAVs & True & 4 & 5M & Highspeed & RandomWaypoint & 10, 25 & 10, 150 \\ 
\hline
\end{tabular}%
}
\label{tab:table1}
\end{table*}

\subsection{Enabling Node Radiation Detection Functionality}

In order to enable some hosts to perform radiation detection in specific zones, some modifications had to be made to the base simulator classes. The \textit{DTNHost} Java class is what each node is created from, and so the changes were made in here for simplicity. After identifying the blast center coordinate through spawning of nodes across the GUI, the zone radii could be defined within the class to be used for determining what zone a host was or was not within. This was done by calculating the distance of the node from the blast center on each movement - but only for radiation detection capable nodes. These were defined with a minimum and maximum range as private static variables at the top of the class for each phase simulation. Appropriate getter and setter methods were created for these functionalities, and their usage is shown in Listing \ref{lst:radiationzonechecking}.

\begin{figure}[!h]
\centering
\begin{minipage}{\linewidth}
\lstset{style=ieeecode, caption={DTNHost changes: creating a message on zone change.}, label={lst:radiationzonechecking}}
\begin{lstlisting}
if (this.address >= rad_detecting_minimum_node) {
    double distanceToBlastCenter = this.location.distance(blastCenter);
    Zone newZone = getZoneForDistance(distanceToBlastCenter);
    if (newZone != currentZone) {
        DTNHost to = getNodeByAddress(drawToAddress(this.address));
        Message m = new Message(this, to, this.getMessageID(), 100);
        m.setResponseSize(0);
        this.createNewMessage(m);
        currentZone = newZone;
    }
}
\end{lstlisting}
\end{minipage}
\end{figure}

A static list that each created node is added to on initialisation is also created, in order to retrieve a DTNHost by its address to define a destination node for each message, demonstrated in Listing \ref{lst:radiationzonechecking} by \textit{getNodeByAddress()}. This code snippet is run within the node's movement function, only for nodes with addresses greater than or equal to the minimum radiation detecting node address. This is because of my node group configuration, with capable nodes defined last. 

\begin{figure}[!h]
\centering
\begin{minipage}{\linewidth}
\lstset{style=ieeecode, caption={DTNHost changes: selecting a relevant destination node.}, label={lst:selectingdestinationaddr}}
\begin{lstlisting}
protected int drawToAddress(int from) {
	    int to;
	    do {
	        to = rad_detecting_minimum_node + random.nextInt(rad_detecting_maximum_node 
            - rad_detecting_minimum_node);
	    } while (from == to);
	    return to;
	}
\end{lstlisting}
\end{minipage}
\end{figure}

Listing \ref{lst:selectingdestinationaddr} shows the method to draw an appropriate destination address from the range of all detection capable nodes, ensuring that a node does not select itself as the recipient. These were selected to simulate information being sent to decision-makers.

\section{Multi-Dimensional Protocol Performance Analysis}

Running the simulations provided insights into the performance of the Epidemic and Prophet protocols under
the varying conditions of the different PNE phases. The following results were obtained for three key performance
metrics: delivery probability, overhead ratio, and latency. Each graph and its corresponding table display
how the protocols performed.

\subsection{Radiation Data Message Delivery Probability}

\begin{figure}[!ht]
    \centering
    \subfloat[Radiation data message delivery probability values]{
        \begin{minipage}[t]{0.38\textwidth}
            \centering
            \renewcommand{\arraystretch}{1.2}
            \begin{tabular}{|c|c|c|}
                \hline
                \rowcolor[HTML]{E0E2FF} 
                \textbf{PNE Phase} & \textbf{Epidemic} & \textbf{PRoPHET} \\
                \hline
                \textbf{1} & 0.9259 & 0.9612 \\
                \textbf{2} & 0.9820 & 0.9777 \\
                \textbf{3} & 0.9971 & 0.9959 \\
                \textbf{4} & 0.9987 & 0.9978 \\
                \hline
            \end{tabular}
        \end{minipage}
    }\hspace{1em}
    \subfloat[Radiation data message delivery probability plot]{
        \begin{minipage}[t]{0.52\textwidth}
            \centering
            \hspace*{-4em} 
            \includegraphics[width=\textwidth]{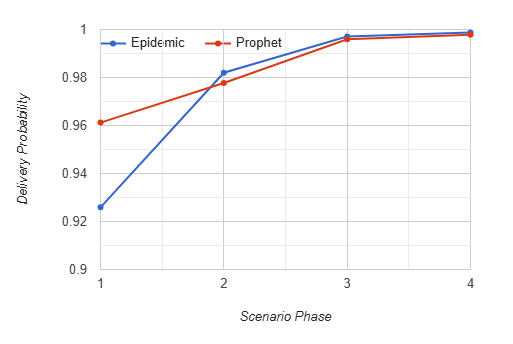}
        \end{minipage}
    }
    \caption{Radiation data message delivery probability results for both protocols across simulation phases.}
    \label{fig:delivery_prob}
\end{figure}

The delivery probability represents the proportion of radiation data messages successfully delivered to their intended recipients. This is crucial for understanding how effective a protocol has been in ensuring message transmission, particularly in a dynamic environment. 

In the initial phase, the network experiences significant challenges due to disrupted movement and minimal radiation detection capable nodes entering the zones. For the Epidemic simulation 378 messages were created, and 412 for PRoPHET. Despite these conditions, Epidemic achieves a delivery probability of 0.9259, slightly lower than PRoPHET at 0.9612. This difference suggests PRoPHET performs better routing decisions with limited nodes. 

Delivery probability increases in Phase 2 (Epidemic 0.982, PRoPHET: 0.9777), with 891 and 897 messages created for Epidemic and PRoPHET respectively. The increase may be due to more nodes moving within the zones because of the decrease in the radiation intensity, leading to more connected nodes and easier communication. Unlike the initial phase, Epidemic slightly outperforms PRoPHET here. This is likely due to the use of the flooding approach, whereas PRoPHET relies on probabilistic routing, and where mobility patterns are chaotic and less predictable, the calculated probabilities may be less effective. 

The delivery probability continues to rise in Phase 3 (Epidemic: 0.9971, PRoPHET: 0.9959). There is a larger time period in this phase, which gives messages more time to be delivered, and PRoPHET more time to build its probability table. Additionally, there are more nodes present within zones, increasing network connectivity. This is also reflected in the messages created, as more radiation detection capable nodes enter the zones, with the Epidemic simulation producing 11872 messages, and PRoPHET's producing 14278. 

Finally, in Phase 4, the delivery probability for both protocols peaks (Epidemic: 0.9987, PRoPHET: 0.9978). This is almost perfect, likely due to the continued reduction in radiation levels and eventual stabilisation of the environment, as well as the greatest length of time for a phase. Despite PRoPHET having a more intelligent decision-making process, it was slightly outperformed by Epidemic in almost all phases. This highlights the advantages of its flooding-based approach in ensuring message delivery. Replicating messages to every node it encounters ensures wide dissemination, which is a contrast to PRoPHET's selective forwarding strategy. Selective forwarding in this instance is less effective as all nodes are unpredictable, and there is no particular set place that a destination node could be. However, the difference in the two protocols is minimal, and Epidemic's higher delivery probability usually comes at the cost of an increased overhead ratio, and these tradeoffs must be considered.

\subsection{Network Overhead}

\begin{figure}[!ht]
    \centering
    \subfloat[Network overhead ratio values]{
        \begin{minipage}[t]{0.38\textwidth}
            \centering
            \renewcommand{\arraystretch}{1.2}
            \begin{tabular}{|c|c|c|}
                \hline
                \rowcolor[HTML]{E0E2FF} 
                \textbf{PNE Phase} & \textbf{Epidemic} & \textbf{PRoPHET} \\ \hline
                \textbf{1} & 118.0914 & 101.5606 \\ \hline
                \textbf{2} & 80.6343 & 70.0582 \\ \hline
                \textbf{3} & 93.792 & 85.6947 \\ \hline
                \textbf{4} & 157.6279 & 136.7805 \\ \hline
            \end{tabular}
        \end{minipage}
    }\hspace{1em}
    \subfloat[Network overhead ratio plot]{
        \begin{minipage}[t]{0.52\textwidth}
            \centering
            \hspace*{-3em} 
            \includegraphics[width=\textwidth]{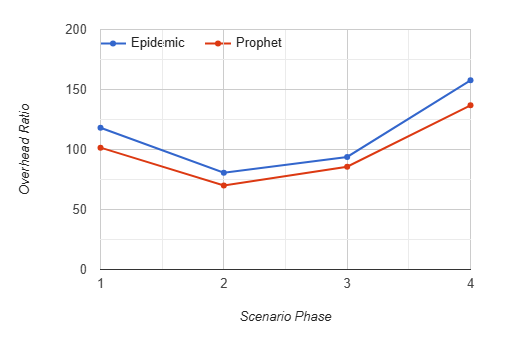}
        \end{minipage}
    }
    \caption{Network overhead ratio results for both protocols across simulation phases.}
    \label{fig:overhead_ratio}
\end{figure}

The overhead ratio measures the ratio of the total number of messages sent in the system to the number of messages successfully delivered. A higher overhead indicates greater network resource consumption. This helps to understand the cost of message replication and overall efficiency of the protocols in each phase. 

In the first phase, both protocols exhibit high overhead ratios (Epidemic: 118.0914, PRoPHET: 101.5606). This is expected due to the initial conditions being more sparse, as limited nodes are within the disaster zones, leaving many nodes unable to communicate directly. More flooding is likely to be needed to find delivery paths, but this also creates numerous redundant copies. PRoPHET's overhead is notably lower, indicating less network congestion. 

In the next phase, the movement of nodes is more structured, but less present inside the zones as people remain in shelter. PRoPHET continues to have a lower overhead ratio (Epidemic: 80.6343, PRoPHET: 70.0582), suggesting better resource utilisation. PRoPHET may also leverage historical encounter data here which could also reduce overhead. 

Phase 3 shows a slight increase in overhead ratio for both protocols (Epidemic: 93.792, PRoPHET: 85.6947). More nodes are able to enter the zones now, and more people will leave their shelter to evacuate since it will have been a significant enough amount of time. Along with the increased length of the phase, these factors result in an increased number of message copies propagated. Epidemic will have more duplicates due to flooding, and PRoPHET will have more opportunities for replication, but will have less of an increase in overhead due to its forwarding strategy, maintaining its comparative efficiency.

In the final phase, there is a significant spike in overhead for both protocols. The extended simulation period means that many messages remain active in the network, leading to more transmission events as nodes encounter each other. Additionally, as radiation levels drop, nodes may traverse larger areas or interact with a greater variety of groups, increasing message propagation. Both overheads are at their highest (Epidemic: 157.6279, PRoPHET: 136.7805), and for Epidemic the network has likely reached saturation, with redundant messages being sent even when nodes already hold copies. PRoPHET's selective approach mitigates this growth in comparison. Across all phases, PRoPHET consistently has a lower overhead ratio compared to Epidemic, thanks to its design focus on resource efficiency. This is at the expense of slightly lower delivery probability.

\subsection{Latency Average of Radiation Data Delivery}

\begin{figure}[!ht]
    \centering
    \subfloat[Latency average values]{
        \begin{minipage}[t]{0.38\textwidth}
            \centering
            \renewcommand{\arraystretch}{1.2}
            \begin{tabular}{|c|c|c|}
                \hline
                \rowcolor[HTML]{E0E2FF} 
                \textbf{PNE Phase} & \textbf{Epidemic} & \textbf{PRoPHET} \\ \hline
                \textbf{1} & 1444.7251 & 1795.6876 \\ \hline
                \textbf{2} & 1193.2686 & 1584.6333 \\ \hline
                \textbf{3} & 1166.6066 & 1927.6702 \\ \hline
                \textbf{4} & 831.8998 & 1336.6147 \\ \hline
            \end{tabular}
        \end{minipage}
    }\hspace{1em}
    \subfloat[Latency average plot]{
        \begin{minipage}[t]{0.52\textwidth}
            \centering
            \hspace*{-3em} 
            \includegraphics[width=\textwidth]{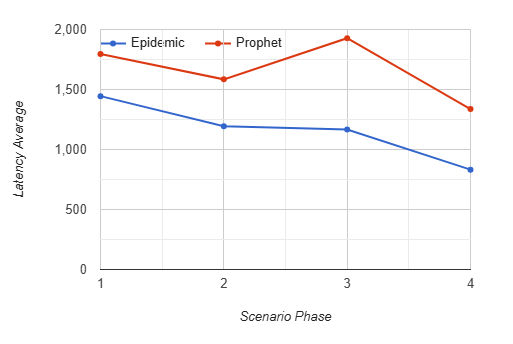}
        \end{minipage}
    }
    \caption{Latency average of radiation data delivery results for both protocols across simulation phases.}
    \label{fig:latency_average}
\end{figure}

Latency measures the average time it takes for a message to be delivered from the source node to the destination node. This is important to evaluate how responsive the network is, particularly in time-sensitive scenarios such as this. Lower latency is preferable as it indicates faster radiation data message delivery.

In phase 1, latency is relatively high (Epidemic: 1444.7251, PRoPHET: 1795.6876). This is likely due to the post-blast chaos, lack of radiation detection capable nodes in the area (and therefore few potential destination nodes), and a more sparse network. Both protocols take longer to find paths, relying on multi-hop routes. Epidemic has a substantially faster delivery than PRoPHET, with its flooding disseminating messages widely and quickly, increasing the likelihood of a message reaching its destination sooner. As the nodes move very unpredictably, especially in phase 1, PRoPHET misses forwarding opportunities as it tries to calculate the best route forwards, despite there probably not being a likely route to find.

Phase 2 shows decreased latency, with Epidemic maintaining its lead (Epidemic: 1193.2686, PRoPHET: 1584:6333). With less nodes allowed in the affected zones, more will gather in closer proximity outside, decreasing latency. This helps Epidemic more as its messages will propagate much faster with nodes in close proximity, compared to PRoPHET which will still be trying to calculate the best route. 

In phase 3, Epidemic dramatically outperforms PRoPHET in terms of latency (Epidemic: 1166.6066, PRoPHET: 1927.6702), achieving its lowest latency of all phases, while PRoPHET experiences its highest. Epidemic's flooding continues to deliver messages faster, as PRoPHET's calculations go to waste with node movement being mostly random. Epidemic ensures a node passes on messages to every other node it encounters, whereas with PRoPHET a node may have to have multiple encounters before it's forwarded - this is a big limitation in this scenario.

Finally, phase 4 shows a greater reduction in latency for both protocols, with Epidemic achieving its lowest latency by far, and PRoPHET narrowing the gap but still lagging behind (Epidemic: 831.8998, PRoPHET: 1336.6147). The most nodes in the simulation are present in this phase, which helps to improve latency times for both protocols as the network is more connected. Epidemic's more aggressive delivery mechanism results in it delivering messages much faster especially when in closer proximity to other nodes. While PRoPHET's efficiency improves over time, it cannot match Epidemic's rapid delivery speed due to its selective forwarding strategy. Epidemic is clearly a better choice for faster message delivery alone.

\section{Discussion and Future Work}

The results of the simulation demonstrated clear performance trade-offs between Epidemic and PRoPHET routine protocols under the created post-nuclear disaster conditions. Epidemic consistently achieved higher message delivery probabilities and did so with lower latency. This performance can be attributed to its flooding-based approach, which proves advantages in highly dynamic and unpredictable mobility environments, which is typical of emergency scenarios. In such cases, rapid and reliable message dissemination is critical and potentially influences life-critical decisions. However, this came at the cost of a higher network overhead ratio due to excessive message copies. 

Although this simulation assumed sufficiently large buffer sizes to absorb this overhead without penalty, practical implementations in resource-constrained environments may not be as tolerant. In these contexts, Epidemic's high network overhead could overload node buffers leading to network congestion, thereby negating its delivery advantage. Conversely, PRoPHET produced slightly lower message delivery probabilities and modestly higher latency, but had much lower network overhead. This trade-off may be acceptable in real-world deployments where bandwidth, buffer capacity, and energy are limited, and where a modest delay in message delivery is permissible.

The simulation findings suggest that Epidemic is better suited for scenarios where message delivery reliability and speed are paramount, provided that node resources are sufficient. On the other hand, PRoPHET is a viable alternative for more resource constrained environments, as long as some message delivery performance can be sacrificed.

Future work could extend this study by incorporating more diverse and realistic node behaviour, such as role-based mobility (e.g., civilians and first responders having more defined movement patterns instead of set areas). Additionally, evaluating the impact of varying buffer sizes, transmission ranges, and message TTLs would offer deeper insight into individual protocol stability. Exploring hybrid or adaptive routing strategies that dynamically switch based on network state could also be beneficial, as no disaster zone is ever the same. For example, the approach proposed in \cite{cognitivecaching}, which uses multi-agent deep reinforcement learning for cognitive caching in mobile social networks, demonstrates the potential of intelligent, context-aware decision making at the network edge. Combining this approach with a low-cost platform such as MODiToNeS \cite{moditones} could facilitate the practical prototyping and validation of these adaptive strategies, and simulating additional environmental hazards and longer-duration scenarios would further improve the realism and applicability of the results.

\bibliographystyle{IEEEtran}
\bibliography{bibliography}

\end{document}